# A Novel Gaussian Based Similarity Measure for Clustering Customer Transactions Using Transaction Sequence Vector

**M.S.B.Phridvi Raj[1], Vangipuram Radhakrishna[2], C.V.Guru Rao[3]**
[1] prudviraj.kits@gmail.com, Department of CSE, Kakatiya Institute of Technology and Science, Warangal, India.
[2] radhakrishna_v@vnrvjiet.in, Department of Information Technology, VNR VJIET, Hyderabad, India.
[3] Principal and Professor, S.R.Engineering College, Warangal, India.

**Abstract.** Clustering Transactions in sequence, temporal and time series databases is achieving an important attention from the database researchers and software industry. Significant research is carried out towards defining and validating the suitability of new similarity measures for sequence, temporal, time series databases which can accurately and efficiently find the similarity between user transactions in the given database to predict the user behavior. The distribution of items present in the transactions contributes to a great extent in finding the degree of similarity between them. This forms the key idea of the proposed similarity measure. The main objective of the research is to first design the efficient similarity measure which essentially considers the distribution of the items in the item set over the entire transaction data set and also considers the commonality of items present in the transactions, which is the major drawback in the Jaccard, Cosine, Euclidean similarity measures. We then carry out the analysis for worst case, the average case and best case situations. The Similarity measure designed is Gaussian based and preserves the properties of Gaussian function. The proposed similarity measure may be used to both cluster and classify the user transactions and predict the user behaviors.

## 1. INTRODUCTION

Clustering Transactions in sequence databases, temporal databases, and time series databases is achieving an important attention from the database researchers and from the perspective of the software industry. The importance for clustering comes from the need for decision making such as classification, prediction. The input to clustering algorithm in databases is usually a set of user transactions with the output being set of clusters of user transactions. One of the important properties of clustering is, all the patterns within a cluster share similar or properties in some sense and patterns in different clusters are dissimilar in corresponding sense. The advantage of clustering w.r.t databases is that each user transaction has a fixed item set with the item set consisting of fixed set of items and do not change frequently. In other words, the item set is static. This eliminates the need of preprocessing the transaction dataset. The motivation of this work comes from our previous research (M.S.B.Phridvi Raj et.al; 2014).

In this paper, we design the similarity measure for clustering the user transactions which has the Gaussian property and considers the distribution of each item from the item set over the entire database of transactions. In case the transactions are arriving as a stream then we can first find the closed frequent item set and apply the similarity measure on the final set of transactions.

In the recent years, clustering data streams has gained lot of research focus in academia and industry (Albert Bifet et.al; 2011, Chang Dong Wang et.al; 2013, Chen Ling et.al; 2012, Shi Zhong; 2005). An approach for handling text data stream is discussed in (Yu Bao Liu; 2008). A similarity measure for clustering and classification of the text which considers the distribution of words is discussed in (Yung-Shen Lin et.al; 2014) which helped us a lot in carrying out the work. A tree based approach for clustering text stream data using the concept of ternary vector is discussed in (M.S.B.Phridvi Raj et.al; 2013).

## 2. PROPOSED MEASURE

The idea for the present similarity measure comes from our previous work (Phridvi Raj et.al; 2014, Chintakindi Srinivas et.al; 2014) considering the feature distribution and commonality which also holds good between the pair of any two transactions. In this work we assume each transaction to be a sequence of 2-tuple elements, the first being count of each item and the later denoting the presence or absence of an item in that transaction say $T_i$. The table.1 denotes the function Φ, and here we use it as a second element in the 2-tuple representation. We define another function called $\Delta$ ($I_{ik}$, $I_{jk}$) which is used to store the difference of count of items w.r.t transactions $T_i$ and $T_j$. The table.1 and table.2 define functions Φ and Δ for the binary and non-binary transaction-item set.

**Table 1.** Function definitions Φ and Δ for transaction item set in binary form

| $I_{ik}$ | $I_{jk}$ | $\Phi(I_{ik}, I_{jk})$ | $\Delta(I_{ik}, I_{jk})$ |
|---|---|---|---|
| 0 | 0 | U | 0 |
| 0 | 1 | 0 | 1 |
| 1 | 0 | 0 | 1 |
| 1 | 1 | 1 | 0 |

**Table 2.** Function definitions Φ and Δ for transaction itemset in non-binary form

| $I_{ik}$ | $I_{jk}$ | $\Phi(I_{ik}, I_{jk})$ | $\Delta(I_{ik}, I_{jk})$ |
|---|---|---|---|
| 0 | 0 | U | 0 |
| 0 | $C_{jk}$ | 0 | $C_{jk}$ |
| $C_{ik}$ | 0 | 0 | $C_{ik}$ |
| $C_{ik}$ | $C_{jk}$ | 1 | $C_{ik}$ - $C_{jk}$ |

### 2.1 Transaction Vector ( $\acute{\Gamma}_i$ )

Let $\acute{\Gamma}_i$ be any transaction with items defined from the item set I = { $I_1$, $I_2$, $I_3$, …..$I_m$} then the transaction vector is a sequence of 2-tuple elements separated by comma (,) and is denoted by each pair of the form ($C_{ik}$, $E_{ik}$) with $C_{ik}$, $E_{ik}$ being count of item k and presence/absence of item k in transaction $T_i$ respectively.

Let $\acute{\Gamma}_1$ = {($C_{11}$, $E_{11}$), ($C_{12}$, $E_{12}$)…….. ($C_{1m}$, $E_{1m}$)} and $\acute{\Gamma}_2$ = {($C_{21}$, $E_{21}$), ($C_{22}$, $E_{22}$)…….. ($C_{2m}$, $E_{2m}$)} be two transaction vectors. Here $C_{ik}$ denote the count of item k in transaction $T_i$ present and $E_{ik}$ denotes presence or absence of an item in transaction $T_i$ . In case we are using binary representation of items without counting then we denote $C_{ik}$ = 1; if $E_{ik}$=1 or we denote $C_{ik}$ = 0 in the case $E_{ik}$ = 0. If we are maintaining count of each item in transaction, then $C_{ik}$ can be any count if $E_{ik}$= 1and $C_{ik}$ = 0 in the case $E_{ik}$=0.

### 2.2 Sequence Vector (SV [$\acute{\Gamma}_i$, $\acute{\Gamma}_j$])

Let $\acute{\Gamma}_i$ and $\acute{\Gamma}_j$ be any two transaction vectors with items defined from the item set I = { $I_1$, $I_2$, $I_3$ …., $I_m$} then the sequence vector over $\acute{\Gamma}_i$ and $\acute{\Gamma}_j$ is defined as SV [$\acute{\Gamma}_i$, $\acute{\Gamma}_j$] = $U_k$ {$T<i, j>_{k}$} = $U_k$ {($\Delta_k^{i,j}$ , $\Phi_k^{i,j}$) } with $U_k$ denoting union of all 2-tuple elements and is represented as

$$SV[\acute{\Gamma}_i, \acute{\Gamma}_j] = [T<i, j>_1 , T<i, j>_2, T<i, j>_3 \ldots\ldots\ldots\ldots. T<i, j>_m]$$

where $T<i, j>_k$ is a 2-tuple denoted by $T<i, j>_k = ((C_{ik} – C_{jk}), \Phi (E_{ik}, E_{jk}))$

Let $\acute{\Gamma}_i$ and $\acute{\Gamma}_j$ be any two transaction vectors with items defined from the item set I = { $I_1$, $I_2$, $I_3$ …..$I_m$} then the sequence vector over $\acute{\Gamma}_i$ and $\acute{\Gamma}_j$ is thus given by

$$SV [\acute{\Gamma}_1, \acute{\Gamma}_2] = [T<1, 2>_1 , T<1, 2>_2, T<1, 3>_3 \ldots\ldots\ldots\ldots. T<1, 2>_m] \quad (1)$$

Where

$$T<1, 2>_1 = ( (C_{11} – C_{21}), \Phi (E_{11}, E_{21}) )$$

$$T<1, 2>_2 = ( (C_{12} – C_{22}), \Phi(E_{12}, E_{22} ) )$$

$$T<1, 2>_3 = ( (C_{13} – C_{23}), \Phi(E_{13}, E_{23} ) )$$

…

$$T<1, 2>_m = ( (C_{1m} – C_{2m}),\Phi(E_{1m}, E_{2m} ) )$$

In general, the Sequence Vector for any two transaction vectors $\acute{\Gamma}$i and $\acute{\Gamma}_j$ is given by

$$SV [T_i, T_j] = U_k \{T<i, j>_1 \} = U_k \{ (\Delta_k^{i,j}, \Phi_k^{i,j}) \} \quad (2)$$

with $U_k$ denoting union of all 2-tuple elements.

Now, we generalize the sequence vector of two transaction vectors to represent SV $[T_i, T_j]$ as

$$SV[T_i, T_j] = \{ T\langle i, j\rangle_1 , T\langle i, j\rangle_2, T\langle i, j\rangle_3 \ldots\ldots\ldots\ldots. T\langle i, j\rangle_m\} \quad (3)$$

where

$$T\langle i, j\rangle_k = (\Delta (I_{ik}, I_{jk}), \Phi (I_{ik}, I_{jk})) \quad (4)$$

with

$\Delta (I_{ik}, I_{jk}) = | I_{ik} | - | I_{jk} |$

$\Phi (I_{ik}, I_{jk})$ is the function on item w.r.t the two transactions $T_i$ and $T_j$

m is the no of items in the item set and k varying from 1 to m.

The sequence vector is a 2 tuple of the form (Δ, Φ) with the elements Δ and Φ. Here Δ contains the difference of the count of two items in both transactions $T_i$ and $T_j$. Here we have the count values of items as 0 or 1. Having defined all the required definitions and terms now we now define our proposed similarity measure given by the equation below

$$TSIM = \frac{(1+S(\alpha,\beta))}{2} \quad (5)$$

Where

$$S(\alpha, \beta) = \frac{\sum_{k=1}^{k=m} \alpha\left(T_{ik}, T_{jk}\right)}{\sum_{k=1}^{k=m} \beta\left(T_{ik}, T_{jk}\right)} \quad (6)$$

$$\alpha\left(T_{ik}, T_{jk}\right) = \begin{cases} 0.5 * [1 + e^{-\gamma^2}] & ; \Phi(I_{ik}, I_{jk}) = 1 ; \Delta (I_{ik}, I_{jk}) = 0 \\ -e^{-\gamma^2} & ; \Phi(I_{ik}, I_{jk}) = 0 ; \Delta (I_{ik}, I_{jk}) = 1 \\ 0 & ; \Phi (I_{ik}, I_{jk}) = U ; \Delta (I_{ik}, I_{jk}) = 0 \end{cases} \quad (7)$$

Where

$$\gamma = \frac{\Delta (I_{ik}, I_{jk})}{\sigma_k}$$

and $\sigma_k$ = standard deviation of feature k in all files of training set. (8)

$$\beta\left(T_{ik}, T_{jk}\right) = \begin{cases} 0 & ; \mathbf{\Phi}(I_{ik}, I_{jk}) = U \\ 1 & ; \mathbf{\Phi}(I_{ik}, I_{jk}) \neq U \end{cases} \quad (9)$$

Here, $T_{ik}$ indicates presence or absence of the $k^{th}$ feature in $i^{th}$ transaction. The values of α and β are used to measure the contribution of each feature in finding similarity.

## 3. VALIDATION OF PROPOSED MEASURE

### 3.1. Best Case Scenario

In the best case situation, all the items may be present in the pair of transactions considered. For the best case situation $T_1$ = {1, 1, 1, 1, 1…….m} and $T_2$ = {1, 1, 1, 1, 1…….m}. Then the sequence vector is denoted by $SV_{12}$ and is represented as $SV_{12}$ = <1, 1, 1, 1, 1……..m>. The value of S (α, β) is computed using eq.7 and eq.9 as shown below

$$S(\alpha,\beta) = \frac{\alpha(T_{i1},T_{j1}) + \alpha(T_{i2},T_{j2}) + \alpha(T_{i3},T_{j3}) + \cdots . + \alpha(T_{im},T_{jm})}{\beta(T_{i1},T_{j1}) + \beta(T_{i2},T_{j2}) + \beta(T_{i3},T_{j3}) + \cdots + \beta(T_{im},T_{jm})}$$

$$= \frac{0.5 * [(1 + e^{-\gamma_1{}^2}) + (1 + e^{-\gamma_2{}^2}) + (1 + e^{-\gamma_3{}^2}) \dots\dots\dots..(1 + e^{-\gamma_m{}^2})]}{(1 + 1 + 1 \dots\dots\dots . \text{mtimes})}$$

$$= \frac{0.5 * (1 + 1 + 1 \dots \text{mtimes}) + 0.5 * (e^{-\gamma_1{}^2} + e^{-\gamma_2{}^2} + e^{-\gamma_3{}^2} \dots . \text{mtimes})}{m}$$

For the best case situation the values of $\sigma_k$ for k = 1 to m, approaches zero. This makes the values of $e^{-\gamma_1{}^2}, e^{-\gamma_2{}^2}, e^{-\gamma_3{}^2} \dots\dots . e^{-\gamma_m{}^2}$ become 1.

This means the above equation reduces to

$$= \frac{0.5 * m + 0.5 * m}{m} = \frac{m}{m} = 1$$

In this case, the similarity measure is

$$\text{TSIM} = \frac{(F + 1)}{(\lambda + 1)} = \frac{(1 + 1)}{(1 + 1)} = 1 \qquad (10)$$

The value of TSIM = 1 indicates that the two text files are most similar to each other.

### 3.2 Worst Case Scenario

The worst case situation occurs when all the items are absent in the transactions considered. This means in the worst case worst case $T_1$ = {0, 0, 0, 0, 0…….m} and $T_2$ = {0, 0, 0, 0, 0…….m}. The Sequence Vector is denoted by $SV_{12}$ and is represented as $SV_{12}$ = <U, U, U...m times>.

The value of S (α, β) is computed using eq.7 and eq.9 as shown below

$$S(\alpha,\beta) = \frac{\alpha(T_{i1},T_{j1}) + \alpha(T_{i2},T_{j2}) + \alpha(T_{i3},T_{j3}) + \cdots . + \alpha(T_{im},T_{jm})}{\beta(T_{i1},T_{j1}) + \beta(T_{i2},T_{j2}) + \beta(T_{i3},T_{j3}) + \cdots + \beta(T_{im},T_{jm})}$$

$$= \frac{U}{U} \text{ (indeterminate situation)}$$

$$= -1 \text{ (so return } -1)$$

In this case, the similarity measure is

$$\mathrm{TSIM} = \frac{(\mathrm{F}+1)}{(\lambda+1)} = \frac{(-1+1)}{(1+1)} = 0 \tag{11}$$

The value of TSIM = 0 indicates that the two text files are least similar to each other or dissimilar w.r.t each other.

### 3.3 Average Case Scenario

In the average case situation, $T_1$ = {1, 0, 1, 0….m times} and $T_2$ = {0, 1, 0, ….m times}. Then the Feature Vector is denoted by $SV_{12}$ and is represented as $SV_{12}$ = <1, 1, 1, 1, 1……..m times>. The value of S $(\alpha, \beta)$ is computed as shown below using eq.7 and eq.9

$$\mathrm{S}\,(\alpha,\beta) = \frac{\alpha(\mathrm{T_{i1}},\mathrm{T_{j1}}) + \alpha(\mathrm{T_{i2}},\mathrm{T_{j2}}) + \alpha(\mathrm{T_{i3}},\mathrm{T_{j3}}) + \cdots. + \alpha(\mathrm{T_{im}},\mathrm{T_{jm}})}{\beta(\mathrm{T_{i1}},\mathrm{T_{j1}}) + \beta(\mathrm{T_{i2}},\mathrm{T_{j2}}) + \beta(\mathrm{T_{i3}},\mathrm{T_{j3}}) + \cdots + \beta(\mathrm{T_{im}},\mathrm{T_{jm}})}$$

$$= \frac{(-e^{-\gamma_1{}^2}) + (-e^{-\gamma_2{}^2}) + (-e^{-\gamma_3{}^2}) \dots\dots\dots..(-e^{-\gamma_m{}^2})}{(1+1+1\dots\dots\dots.\mathrm{mtimes})}$$

$$= \frac{-(e^{-\gamma_1{}^2} + e^{-\gamma_2{}^2} + e^{-\gamma_3{}^2} \dots.\mathrm{mtimes})}{\mathrm{m}}$$

Assuming the exponent values all the same, the above equation reduces to

$$= \frac{-\mathrm{m}e^{-\gamma^2}}{\mathrm{m}} = -e^{-\gamma^2} \tag{12}$$

Case 1: $\gamma = 0$.

The value for similarity measure denoted by S is now given by

$$\mathrm{TSIM} = \frac{(1-e^{-\gamma^2})}{(1+1)} = \frac{(1-1)}{(1+1)} = 0 \tag{13}$$

Case 2: $\gamma \neq \infty$**.** Practically it is not infinite.

Then the value for S is

$$\mathrm{TSIM} = \frac{(1-e^{-\gamma^2})}{(1+1)} = = 0.5 * \left(1-e^{-\gamma^2}\right) \tag{14}$$

## 4. CASE STUDY

Consider the transactions with the following items as in Table.3. The Table.4 below shows the Binary representation of the transaction-item matrix. The entire computation is shown for each pair of transactions as shown below. The value of $\lambda$ is assumed as 1 for the purpose of biasing the similarity measure. Here $\Sigma\alpha$ and $\Sigma\beta$ indicates Numerator and Denominator of the function S $(\alpha, \beta)$ respectively.

**Table 3. User transactions with items**

| | FREQUENT ITEMS |
|---|---|
| T1 | { BREAD, BUTTER, JAM} |
| T2 | { JAM,COFFEE,MILK } |
| T3 | {BUTTER , JAM ,COFFEE , MILK } |
| T4 | {BREAD , BUTTER ,JAM , MILK } |
| T5 | {JAM , COFFEE } |
| T6 | { BREAD , BUTTER , MILK } |
| T7 | { BREAD , BUTTER , COFFEE } |
| T8 | {BUTTER , COFFEE } |
| T9 | { BUTTER ,JAM , MILK } |

**Table 4. Transaction-Itemset matrix in binary form**

| | bread | butter | jam | coffee | milk |
|---|---|---|---|---|---|
| T1 | 1 | 1 | 1 | 0 | 0 |
| T2 | 0 | 0 | 1 | 1 | 1 |
| T3 | 0 | 1 | 1 | 1 | 1 |
| T4 | 1 | 1 | 1 | 0 | 1 |
| T5 | 0 | 0 | 1 | 1 | 0 |
| T6 | 1 | 1 | 0 | 0 | 1 |
| T7 | 1 | 1 | 0 | 1 | 0 |
| T8 | 0 | 1 | 0 | 1 | 0 |
| T9 | 0 | 1 | 1 | 0 | 1 |

**4.1 computations**

<T1-T2>: < (1, 0), (1, 0), (0, 1), (1, 0), (1, 0)>

$\Sigma\alpha$ = - 0.02732 -0.00584+1-0.02732 -0.02732 = 0.9122
$\Sigma\beta$ = 5
TSIM = (0.18244+1)/ (1+1) = 0.59122

<T1-T3>: < (1, 0), (0, 1), (0, 1), (1, 0), (1, 0)>

$\Sigma\alpha$ = -0.02732 + 1+1-0.02732-0.02732 =1.91804
$\Sigma\beta$ = 5
TSIM = 0.691804

<T1-T4>: < (0, 1), (0, 1), (0, 1), (0, U), (1, 0)>

$\Sigma\alpha$ = 1 +1+1+0-0.0273 = 2.97268
$\Sigma\beta$ = 4
TSIM= (0.74317+1)/ (1+1) = 0.871585

<T1-T5>: < (1, 0), (1, 0), (0, 1), (1, 0), (0, U)>

$\Sigma\alpha$ = -0.0273-0.00584+1-0.02732+0=0.92704
$\Sigma\beta$ = 4
TSIM= (0.23176+1)/ (1+1) =0.61588

<T1-T6>: < (0, 1), (0, 1), (1, 0), (0, U), (1, 0)>

$\Sigma\alpha$ =1+1-0.01832+0-0.02732=1.95436
$\Sigma\beta$ = 4
TSIM= (0.48859+1)/2=0.744295

<T1-T7>: < (0, 1), (0, 1), (1, 0), (1, 0), (0, U)>

$\Sigma\alpha$= 1 +1-0.0183-0.0273+0=1.95436
$\Sigma\beta$ = 4
TSIM= (0.48859+1)/2 = 0.744295

<T1-T8> = <(1,0), (0,1),(1,0), (1,0), (0,U)>

$\Sigma\alpha$= -0.0273+1-0.0183-0.02732+0=0.92704
$\Sigma\beta$ = 4
TSIM= (0.23716+1)/2 = 0.61588

<T1-T9> =<(1,0), (0,1), (0,1), (0,U), (1,0)>

$\Sigma\alpha$= -0.0273+1+1+0-0.0273=1.94536
$\Sigma\beta$ = 4
TSIM= (0.48634+1)/2 = 0.74317

<T2-T3>= <(0,U), (1,0), (0,1), (0,1), (0,1)>

$\Sigma\alpha$ = 0-0.00584+1+1+1=2.9213
$\Sigma\beta$ = 4
TSIM = 0.8651625

<T2-T4>= <(1,0), (1,0), (0,1), (1,0), (0,1)>

$\Sigma\alpha$ =-0.0273-0.00584+1-0.02732+1=1.8940
$\Sigma\beta$ = 5
TSIM = 0.6894

<T2-T5> = < (0, U), (0, U), (1, 0), (1, 0), (0, 1)>

$\Sigma\alpha$= 0+0-0.0183-0.02732+1=0.9271
$\Sigma\beta$ = 3
TSIM = 0.6545

<T2-T6> = <(1,0), (1,0), (1,0), (1,0), (0,1)>

$\Sigma\alpha$ =-0.0273-0.00584-0.01830-.02732+1=0.9486
$\Sigma\beta$= 5
TSIM = 0.59486

<T2-T7> = < (1,0), (1,0), (1,0), (0,1), (1,0)>

$\Sigma\alpha$ =-0.0273-0.00584-0.01830+1-0.02732=0.9486
$\Sigma\beta$= 5
TSIM = 0.59486

<T2-T8> = < (0, U), (1, 0), (1, 0), (0, 1), (1, 0)>

$\Sigma\alpha$ = 0-0.00584-0.01830+1-0.02732=0.9213
$\Sigma\beta$ = 4
TSIM = 0.6152

<T2-T9> = <(0,U), (1,0), (0,1), (1,0), (0,1)>

Σα = 0-0.00584+1-0.02732+1=1.9213
Σβ = 4
TSIM = 0.7402

<T3-T4> = <(1,0), (0,1), (0,1), (1,0), (0,1)>

Σα = -0.0273+1+1-0.02732+1= 2.8940
Σβ = 5
TSIM= 0.7894

<T3-T5> = <(0,U), (1,0), (0,1), (0,1), (1,0)>

Σα = 0-0.00584+1+1-0.02732=1.9213
Σβ = 4
TSIM= 0.7402

<T3-T6> = <(1,0), (0,1), (1,0), (1,0), (0,1)>
Σα= -0.0273+1-0.01832-0.02732+1=1.8940
Σβ = 5
TSIM= 0.6894

<T3-T7> = <(1,0), (0,1), (1,0), (0,1), (1,0)>

Σα= -0.0273+1-0.01832+1-0.02732=1.8940
Σβ = 5
TSIM= 0.6894

<T3-T8> = <(0,U), (0,1), (1,0), (0,1), (1,0)>

Σα = 0+1-0.01832+1-0.02732 = 1.9213
Σβ = 4
TSIM= 0.7402

<T3-T9> = <(0,U), (0,1), (0,1), (1,0), (0,1)>

Σα = 0+1+1-0.02732+1=2.9213
Σβ = 4
TSIM= 0.8652

<T4-T5> = <(1,0), (1,0), (0,1), (1,0), (1,0)>

Σα =-0.0273-0.00584+1-0.02732-0.02732= 0.8940
Σβ= 5
TSIM= 0.5894

<T4-T6> = <(0,1), (0,1), (1,0), (0,U), (0,1)>

Σα= 1+1-0.01832+0+1= 2.9213
Σβ= 4
TSIM= 0.8652

<T4-T7> = <(0,1), (0,1), (1,0), (1,0), (1,0)>

$\Sigma\alpha$ = 1+1-0.01832-0.02732-0.02732= 1.8940
$\Sigma\beta$ = 5
TSIM= 0.6894

<T4-T8> = <(1,0), (0,1), (1,0), (1,0),(1,0)>

$\Sigma\alpha$=-0.0273+1-0.01832-0.02732-0.02732=0.8940
$\Sigma\beta$ = 5
TSIM= 0.5894

<T4-T9> = < (1, 0), (0, 1), (0, 1), (0, U), (0, 1)>

$\Sigma\alpha$ = -0.0273+1+1+0+1= 2.9213
$\Sigma\beta$= 4
TSIM= 0.8652

<T5-T6> = <(1,0), (1,0), (1,0), (1,0), (1,0)>

$\Sigma\alpha$ =-0.0273-0.00584-0.01832-0.02732-0.02732 = -0.0514
$\Sigma\beta$= 5
TSIM= 0.4949

<T5-T7> = <(1,0), (1,0), (1,0), (0,1), (0,U)>

$\Sigma\alpha$ = -0.0273-0.00584-0.01832+1+0 = 0.9213
$\Sigma\beta$ = 4
TSIM= 0.6152

<T5-T8> = <(0,U), (1,0), (1,0), (0,1), (0,U)>
$\Sigma\alpha$ = 0-0.00584-0.01832+1+0=0.9486
$\Sigma\beta$ = 3
TSIM= 0.6581

<T5-T9> = <(0,U), (1,0), (0,1), (1,0), (1,0)>

$\Sigma\alpha$ = 0-0.00584+1-0.02732-0.02732 = 0.9213
$\Sigma\beta$= 4
TSIM= 0.6152

<T6-T7> = <(0,1), (0,1), (0,U), (1,0), (1,0)>

$\Sigma\alpha$ = 1+1+0-0.02732-0.02732=1.9123
$\Sigma\beta$= 4
TSIM= 0.7390

<T6-T8> = <(1,0), (0,1), (0,U), (1,0), (1,0)>

$\Sigma\alpha$ = -0.0273+1+0-0.02732-0.02732 = 0.9123
$\Sigma\beta$ = 4
TSIM= 0.6140

<T6-T9> = <(1,0), (0,1), (1,0), (0,U), (0,1)>

$\Sigma\alpha$= -0.0273+1-0.01832+0+1= 1.9213
$\Sigma\beta$ = 4
TSIM= 0.7402

<T7-T8> = < (1, 0), (0, 1), (0, U), (0, 1), (0, U)>

$\Sigma\alpha$ = -0.0273+1+0+1+0=1.9396
$\Sigma\beta$ = 3
SIM= 0.8233

<T7-T9> = <(1,0), (0,1), (1,0), (1,0), (1,0)>

$\Sigma\alpha$ = -0.0273+1-0.01832-0.02732-0.02732= 0.8940
$\Sigma\beta$ = 5
TSIM= 0.5894

<T8-T9> = <(0,U), (0,1), (1,0), (1,0), (1,0)>

$\Sigma\alpha$ = 0+1-0.01832-0.02732-0.02732=0.9213
$\Sigma\beta$= 4
SIM= 0.6152

The table.5 below shows the similarity value for each transaction pair called similarity matrix. As the similarity values of the matrix are symmetric, we only show the upper triangular element values in the table.5 depicting similarity matrix.

**Table 5. Similarity Matrix Showing Upper triangular values**

| | T1 | T2 | T3 | T4 | T5 | T6 | T7 | T8 | T9 |
|---|---|---|---|---|---|---|---|---|---|
| T1 | - | 0.59122 | 0.6918 | 0.8715 | 0.6158 | 0.7442 | 0.7442 | 0.6158 | 0.7431 |
| T2 | - | - | 0.8651 | 0.6894 | 0.6545 | 0.5948 | 0.5948 | 0.6152 | 0.7402 |
| T3 | - | - | - | 0.7894 | 0.7402 | 0.6894 | 0.6894 | 0.7402 | 0.8652 |
| T4 | - | - | - | - | 0.5894 | 0.8652 | 0.6894 | 0.5894 | 0.8652 |
| T5 | - | - | - | - | - | 0.4949 | 0.6152 | 0.6581 | 0.6152 |
| T6 | - | - | - | - | - | - | 0.7390 | 0.6140 | 0.7402 |
| T7 | - | - | - | - | - | - | - | 0.8233 | 0.5894 |
| T8 | - | - | - | - | - | - | - | - | 0.6152 |
| T9 | - | - | - | - | - | - | - | - | - |

The final set of Clusters formed after applying Clustering algorithm (Vangipuram et.al; 2014, C.Srinivas et.al; 2013, Phridviraj et.al; 2014) is

Cluster-1: { $T_1$, $T_2$, $T_3$, $T_4$, $T_6$, $T_9$ }

Cluster-2: { $T_7$, $T_8$ }

Cluster-3: { $T_5$ }

## 5. CONCLUSIONS

The objective of this research is to propose a new similarity measure which considers the distribution of the items of the transaction over the entire transaction dataset and can be used for clustering and classification of transactions and also the users based on the transactions carried out. This helps in predicting user behaviors in advance. In this paper, we design and define a novel similarity measure which can be used to cluster the user transactions. The similarity measure is analyzed for worst case, average case and best case situations. To extend the clustering process to data stream of transactions we may use the algorithm defined in (M.S.B.PhridviRaj et.al; 2014). In future we may extend the research to handle the data streams and evaluate the suitability of proposed similarity measure to perform the classification.